\documentclass[twocolumn]{aastex631}

\usepackage{amsmath}

\usepackage{savesym}
\usepackage{gensymb}
\savesymbol{tablenum}
\usepackage{siunitx}
\usepackage{hyperref}
\restoresymbol{SIX}{tablenum}
\submitjournal{PSJ; Received February 9, 2022; Revised April 28, 2022; Accepted May 5, 2022}

\shorttitle{Moderately High Obliquity Promotes Biospheric Oxygenation}
\shortauthors{Barnett and Olson}
\graphicspath{{./}{figures/}}

\begin{document}

\title{Moderately High Obliquity Promotes Biospheric Oxygenation}

\correspondingauthor{Megan N. Barnett}
\email{meganbarnett@uchicago.edu}

\author{Megan N. Barnett}
\affiliation{Department of the Geophysical Sciences, University of Chicago, USA}
\author{Stephanie L. Olson}
\affiliation{Department of Earth, Atmospheric, and Planetary Sciences, Purdue University, USA}





\begin{abstract}
Planetary obliquity is a first order control on planetary climate and seasonal contrast, which has a number of cascading consequences for life. How moderately high obliquity (obliquities greater than Earth's current obliquity up to 45$\degree$) affects a planet's surface physically has been studied previously, but we lack an understanding of how marine life will respond to these conditions. We couple the ROCKE-3D general circulation model to the cGENIE 3D biogeochemical model to simulate the ocean biosphere's response to various planetary obliquities, bioessential nutrient inventories, and biospheric structure. We find that the net rate of photosynthesis increased by 35$\%$ and sea-to-air flux of biogenic oxygen doubled between the 0$\degree$ and 45$\degree$ obliquity scenarios, which is an equivalent response to doubling bioessential nutrients. Our results suggest that moderately high-obliquity planets have higher potential for biospheric oxygenation than their low-obliquity counterparts and that life on moderately high-obliquity habitable planets may be easier to detect with next generation telescopes. These moderately high-obliquity habitable planets may also be more conducive to the evolution of complex life.
\end{abstract}
\keywords{Astrobiology (74) --- Biosignatures (2018) --- Exoplanets(498) --- Habitable Planets(695) --- Ocean-atmosphere interactions(1150)}

\section{Introduction}

\begin{figure*}
    \centering
    \includegraphics[width=\textwidth]{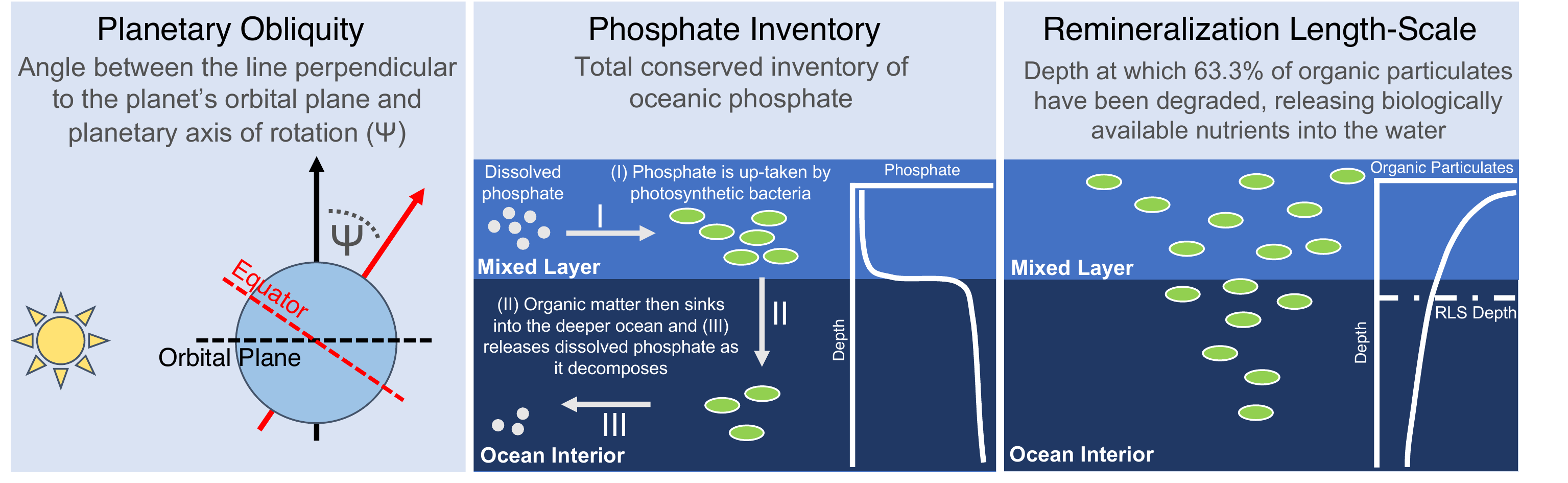}
    \caption{Depiction and definitions of parameters of interest varied in models in this paper. The white dash-dotted horizontal line in right-most panel indicates the depth of the remineralization length-scale (RLS) in the schematic.}
    \label{fig:fig1}
\end{figure*}

Habitable exoplanets will differ from the Earth in several ways that may affect biogeochemical cycles, biospheric productivity, and biospheric detectability. One potential characteristic that might differ between Earth and exoplanets is planetary obliquity, or the angle between the planet’s axis of rotation and orbital axis (Figure \ref{fig:fig1}, left-most panel). Planetary obliquity can significantly affect planetary climate. Planets with larger obliquities have larger amplitude seasonal variations \citep[e.g.][]{williams_habitable_1997, spiegel_habitable_2009, armstrong_effects_2014, nowajewski_atmospheric_2018, guendelman_atmospheric_2019} due to differences in seasonal stellar energy distribution at the planetary surface with increasing obliquity. Moderately high-obliquity planets also experience a more uniform distribution of stellar energy across the planetary surface on annual average, leading these planets to be warmer on average than their low obliquity counterparts \citep[e.g.][]{linsenmeier_climate_2015, wang_effects_2016, nowajewski_atmospheric_2018, kang_mechanisms_2019, guendelman_atmospheric_2019, palubski_habitability_2020, komacek_constraining_2021}.

There is a wide range of obliquities amongst the terrestrial and giant planets in our solar system, ranging from slight to extreme. The terrestrial planets span a wide range of obliquities, with Mercury and Venus having obliquities very close to 0$\degree$ (0.1$\degree$ and 177.4$\degree$ respectively, with Venus orbiting retrograde) and the Earth and Mars currently having moderate obliquities of 23.4$\degree$ and 25$\degree$, respectively. Earth’s obliquity is believed to have varied within the moderate range of 22.1$\degree$ to 24.5$\degree$ for at least the past 400 Myr \citep[][]{williams_history_1993}, while Mars' obliquity may have varied between 0.2$\degree$ to $\sim$60$\degree$ over the last 10-45 Myr \citep[][]{bills_rigid_1990, laskar_chaotic_1993, laskar_long_2004}. Among the outer solar system giant planets, Jupiter, Saturn, and Neptune have obliquities of 3$\degree$, 26.7$\degree$, and 29.6$\degree$, while only Uranus has a much higher obliquity of 98$\degree$. These vastly different obliquities and their evolution in our own solar system suggest that exoplanet obliquities likely vary widely as well, though there are currently very few constraints on these measurements for individual exoplanets \citep[][]{bryan_obliquity_2020}.

Planetary obliquity does not just affect the planet's climate. \citet{olson_oceanographic_2020}  used ROCKE-3D \citep{way_resolving_2017}, a state-of-the-art ocean-atmosphere general circulation model (GCM), to investigate the effect of planetary obliquity on ocean dynamics and marine habitats. They found that increasing planetary obliquity from 0$\degree$ to 45$\degree$ yielded warmer climates and increased seasonal variability of the mixed layer depth (the depth to which the water column is homogeneous due to turbulence and mixing). \citet{olson_oceanographic_2020} argued that ocean circulation patterns changes due to increased planetary obliquity could yield greater nutrient recycling, greater biosphere productivity, and greater biosignature accumulation. However, the GCM used by \citet{olson_oceanographic_2020} lacked explicit representation of life. We address that gap here.

In this work, we couple winds from a general circulation model (ROCKE-3D) with a marine biogeochemical model (cGENIE) to explicitly simulate the response of an Earth-like biosphere to moderately high planetary obliquity (\hyperref[sec:Methods]{Section 2}). \hyperref[sec:Results]{Section 3} describes the biospheric response to various planetary obliquities, phosphate inventories, and biospheric structure. We then discuss our results and their implications for planetary oxygenation, biological evolution, and exoplanet life detection in \hyperref[sec:Discussion]{Section 4}. These findings are summarized in \hyperref[sec:Conclusions]{Section 5}.

\section{Methods}
\label{sec:Methods}
We investigate the response of ocean life to planetary obliquity using the ROCKE-3D GCM leveraged by \citet{olson_oceanographic_2020} to force cGENIE, a 3D biogeochemical model originally developed by \citet{ridgwell_marine_2007}. cGENIE couples a 3D marine biogeochemistry model to a 3D ocean circulation model with dynamic-thermodynamic sea-ice, all under an energy-moisture balance model (EMBM) with bulk atmospheric chemistry. While cGENIE has historically been used to simulate the geologically recent past and near future, it has since been extended to include a full treatment of methane-based atmospheric chemistry, ocean/atmosphere methane cycling, and methane-based metabolisms to enable the simulation of low oxygen biospheres \citep[][]{olson_quantifying_2013, olson_limited_2016, reinhard_oceanic_2020}.

Our experiments use the same topography and bathymetry as the original ROCKE-3D experiments from \citet{olson_oceanographic_2020}, and we prescribe steady-state surface wind output from their experiments to drive ocean circulation within cGENIE. Prescribing steady-state winds greatly reduces the computational cost of our simulations, allowing us to explore a larger swath of parameter space in our simulations. Using steady-state, rather than time-variable, winds is acceptable because wind-driven upwelling was only weakly sensitive to obliquity in the simulations from \citet{olson_oceanographic_2020}. In contrast, the wind-mixed layer depth varied dramatically between seasons in moderately high obliquity scenarios due differences in thermal stratification. It is this mixed layer seasonality that \citet{olson_oceanographic_2020} argued will have the greatest affect on biogeochemical cycles on moderately high-obliquity worlds, and we include this effect in our simulations.

We prescribe an atmosphere with fixed 1$\%$ present atmospheric levels (PAL) of oxygen, similar to Proterozoic earth \citep[e.g.][]{lyons_rise_2014}. Prescribing atmospheric oxygen at low levels allows us to explore oxygen dynamics on on planets that have have not yet experienced a large-scale oxygenation despite the presence of oxygenic photosynthesis, like early Earth where biological oxygen production preceded atmospheric oxygenation by several hundred million years \citep[e.g.][]{planavsky_evidence_2014}. 

Unless otherwise specified, all other model parameters are set at their pre-industrial Earth values (e.g. solar insolation is set at 1368 W/m$^{-2}$ and pCO$_{2}$ is set at 278 ppmv). This includes eccentricity, which is equal to 0.0167, with perihelion occurring during the northern hemisphere winter and introducing slight seasonality even in our 0$\degree$ obliquity scenario.

We consider four different planetary obliquity values, with two low planetary obliquities less than Earth's current obliquity (0$\degree$ and 15$\degree$), and two moderately high planetary obliquities greater than Earth's current obliquity (30$\degree$ and 45$\degree$). We choose these obliquity values to sample the same obliquity space as \citet{olson_oceanographic_2020}, whose experiments provided necessary wind field inputs for our experiments.

In addition to planetary obliquity, we also test the effect of varying two key biosphere parameters. The first of these parameters is the total ocean phosphate inventory. Phosphate (PO$_{4}^{3-}$) is the carrier molecule for phosphorus, an essential element that limits biological activity on Earth over geologic timescales \citep[][]{tyrrell_relative_1999,reinhard_evolution_2017}. As phosphate is a critical nutrient for life, it is important to understand how a range of phosphate inventories would affect life on potential habitable exoplanets.

The second parameter varied in our simulations is the remineralization length-scale (RLS). Organic carbon decays according to the remineralization length-scale within the cGENIE model, implemented as an e-folding depth at which 63.3$\%$ of particulate organic carbon (POC) and associated nutrients have been recycled. We vary this length scale to characterize how the depth-distribution of POC remineralization affects nutrient cycles. 

We run simulations where these three variables (phosphate inventory, remineralization length-scale, and obliquity) are both varied independently and co-varied, for a total of 80 distinct model configurations. Unless otherwise stated, all other model parameters are set to their present-day Earth values. We run the simulations until steady state is achieved (10,000 years). Then, we calculate both annual and seasonal averages for each model output parameter from the final year of model evolution. 

\section{Results}
\label{sec:Results}
\begin{figure*}
    \centering
    \includegraphics[width=\textwidth]{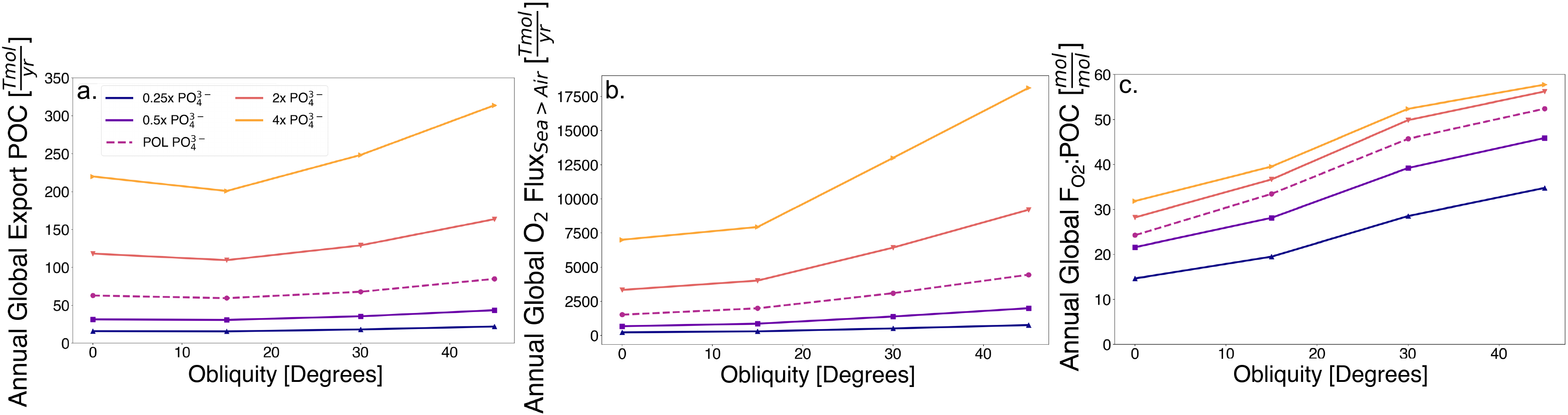}
    \caption{Export particulate organic carbon (POC, panel a), oxygen flux (F$_{O_{2}}$, panel b), and oxygen flux per unit organic carbon (F$_{O_{2}}$:POC, panel c)  from varying phosphate levels at different obliquity levels. The POL label stands for present-day ocean levels, and indicates the present-day Earth phosphate inventory scenario. Other scenarios are labeled with their modification factor times present-day ocean levels. Remineralization length-scale is tuned to match chemical profiles from the Earth's present-day ocean and present day obliquity \citep[][]{ridgwell_marine_2007}.}
    \label{fig:fig2}
\end{figure*}

\subsection{Planetary Obliquity Scenarios}

We first consider the results from a range of obliquity scenarios (0$\degree$ to 45$\degree$) with present-day Earth ocean phosphate inventory and remineralization length-scale to determine planetary obliquity's effect on export POC, the flux of organic carbon produced by photosynthesis that evades remineralization in surface waters and sinks to the deeper ocean--leaving behind oxygen. We additionally quantify the response of sea-to-air oxygen flux resulting from photosynthesis. In these obliquity scenarios (ranging from 0$\degree$ to 45$\degree$), annual total global export POC generally increases with increasing obliquity, with export POC increasing by 35$\%$ between the 0$\degree$ and 45$\degree$ obliquity scenarios. However, export POC reaches a minimum at 15$\degree$ obliquity (Figure \ref{fig:fig2}a, magenta line), decreasing by 5$\%$ between the 0$\degree$ and 15$\degree$ obliquity scenarios.

We also see that global oxygen sea-to-air flux (F$_{O_{2}}$) increases with obliquity. There is a two-fold increase in F$_{O_{2}}$ between the 0$\degree$ obliquity and 45$\degree$ obliquity scenarios (Figure \ref{fig:fig2}b, magenta line). While a portion of this increase in F$_{O_{2}}$ from the ocean to the atmosphere is a consequence of increased photosynthesis, as evidenced by the increase in export POC values with increasing obliquity, the ratio of F$_{O_{2}}$ to POC also increases with higher obliquity (Figure \ref{fig:fig2}c, magenta line). This increase in F$_{O_{2}}$:POC in moderately high-obliquity scenarios suggests that physical effects caused by moderately high obliquity also contribute to the increase in F$_{O_{2}}$ with higher obliquity.

\subsection{Biospheric Sensitivity to Phosphate}
We then simulate oceans with a range of phosphate inventories and planetary obliquity values. Increasing phosphate inventory in our simulations has an $\sim$ 1:1 effect on increasing export POC (Figure \ref{fig:fig2}a). The phosphate inventory also affects the biospheric response to obliquity. The biosphere is relatively insensitive to obliquity when phosphate is scarce (0.25 - 0.5x POL), but export POC becomes increasingly sensitive to obliquity under phosphate replete conditions (2-4x POL).

This same trend is seen in F$_{O_{2}}$ values as well (Figure \ref{fig:fig2}b). At low phosphate inventory levels, F$_{O_{2}}$ is relatively unaffected by obliquity. However, F$_{O_{2}}$ increases more strongly with obliquity when phosphate is more abundant. At high phosphate inventory (2-4x POL), increasing obliquity from 0$\degree$ to 45$\degree$ yields greater that 2x F$_{O_{2}}$, an effect similar to doubling the phosphate inventory. 

The ratio of F$_{O_{2}}$ from sea-to-air to unit export POC experiences a similar two-fold increase with obliquity increase from 0$\degree$ to 45$\degree$ as is caused by a doubling of phosphate inventory (Figure \ref{fig:fig2}c). Unlike for F$_{O_{2}}$ values, scenarios at every phosphate inventory value experienced a doubling in F$_{O_{2}}$:POC ratio with an obliquity increase from 0$\degree$ to 45$\degree$.

\subsection{Biospheric Sensitivity to Remineralization Length-Scale}
As POC sinks through the water column, nutrients are made re-available for metabolic reactions such as photosynthesis through remineralization at depth and transport back to the photic zone. Remineralization is the first step in this process, and occurs throughout the water column. The remineralization length-scale affects the vertical distribution of nutrients in the water column and the timescale for their return to the surface. When remineralization length-scales are longer, nutrient containing particles are able to settle deeper into the water column on average before they are processed and made available for transport back to the surface, leading to greater accumulation of nutrients at depth.

\begin{figure*}
    \centering
    \includegraphics[width=\textwidth]{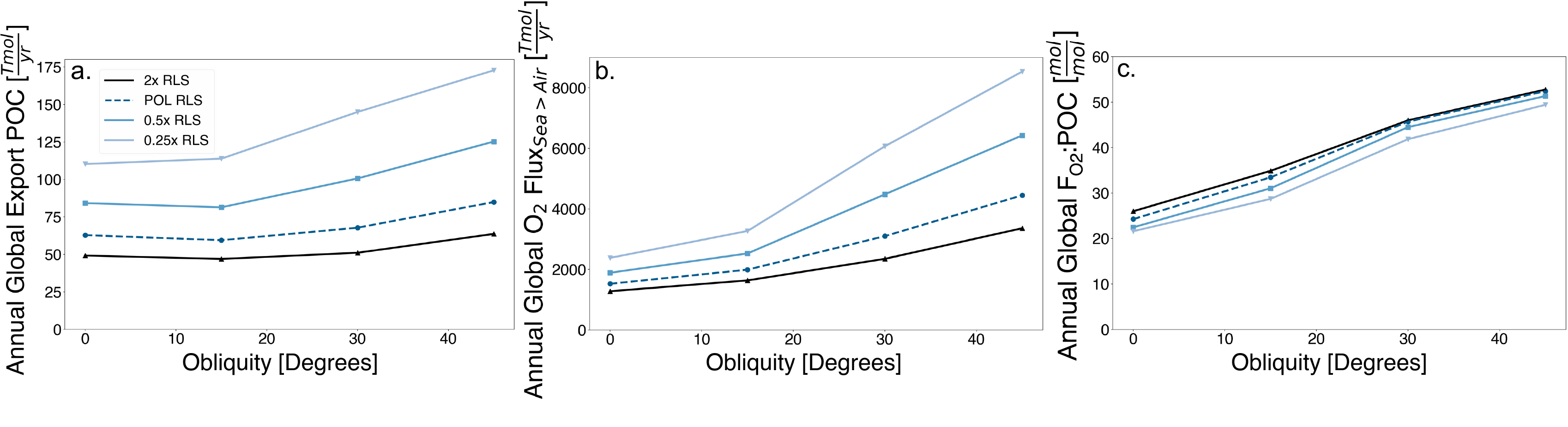}
    \caption{Export particulate organic carbon (POC, panel a), oxygen flux (F$_{O_{2}}$, panel b), and oxygen flux per unit organic carbon (F$_{O_{2}}$:POC, panel c) for scenarios with various remineralization length-scales (each scenario is labeled by the modifying factor times remineralization length-scale) at different obliquity levels. Phosphate inventory in each of these experiments is set to POL.}
    \label{fig:fig3}
\end{figure*}

Simulations with shallower remineralization length scales displayed increased POC export and biosphere activity (Figure \ref{fig:fig3}a). The increase in POC export with shallower remineralization occurs because nutrients do not settle as far into the deep ocean and nutrients are more readily returned to the surface via wind-driven upwelling. 

We find that export POC only slightly increases with obliquity for experiments with longer remineralization length scales. However, export POC increases with higher obliquity for scenarios with shorter remineralization length-scales (0.25 - 0.5x present ocean levels, or POL, Figure \ref{fig:fig3}a). F$_{O_{2}}$ increases by $\sim$3-4x with increasing obliquity and decreasing remineralization length-scale in all our sensitivity experiments at POL phosphate inventory (Figure \ref{fig:fig3}b). Moderately high obliquity is a stronger driver of F$_{O_{2}}$ increase than remineralization length-scale decrease, as halving remineralization length-scale causes an increase in F$_{O_{2}}$ of approximately 50$\%$, while increasing obliquity from 0$\degree$ to 45$\degree$ results in greater than double F$_{O_{2}}$ ($\geq$ 100$\%$ increase).

The ratio of F$_{O_{2}}$ to export POC increases with increasing obliquity (Figure \ref{fig:fig3}c) for all simulated remineralization length-scales. This result also supports our assertion that physical changes caused by increasing obliquity are partially responsible for the observed increase in F$_{O_{2}}$. 

Additionally, we find that while F$_{O_{2}}$ and POC alone decrease with deeper remineralization length-scales, F$_{O_{2}}$:POC ratio increases with deeper remineralization length-scales. F$_{O_{2}}$ and export POC decrease with increasing remineralization length-scale as when the remineralization length-scale is deeper, nutrients sink to deeper levels in the water column and decrease the amount of nutrients available for the surface biosphere. A decrease in photosynthesis due to a reduction of nutrients then leads to a decrease in oxygen production, and by extension F$_{O_{2}}$ decrease. However, when we look at the F$_{O_{2}}$:POC ratio, we see this ratio increase with deeper remineralization depths. Although both F$_{O_{2}}$ and POC decrease with increasing RLS, F$_{O_{2}}$:POC increases because deeper remineralization results in less near surface oxygen consumption.

\begin{figure*}
    \centering
    \includegraphics[width=\textwidth]{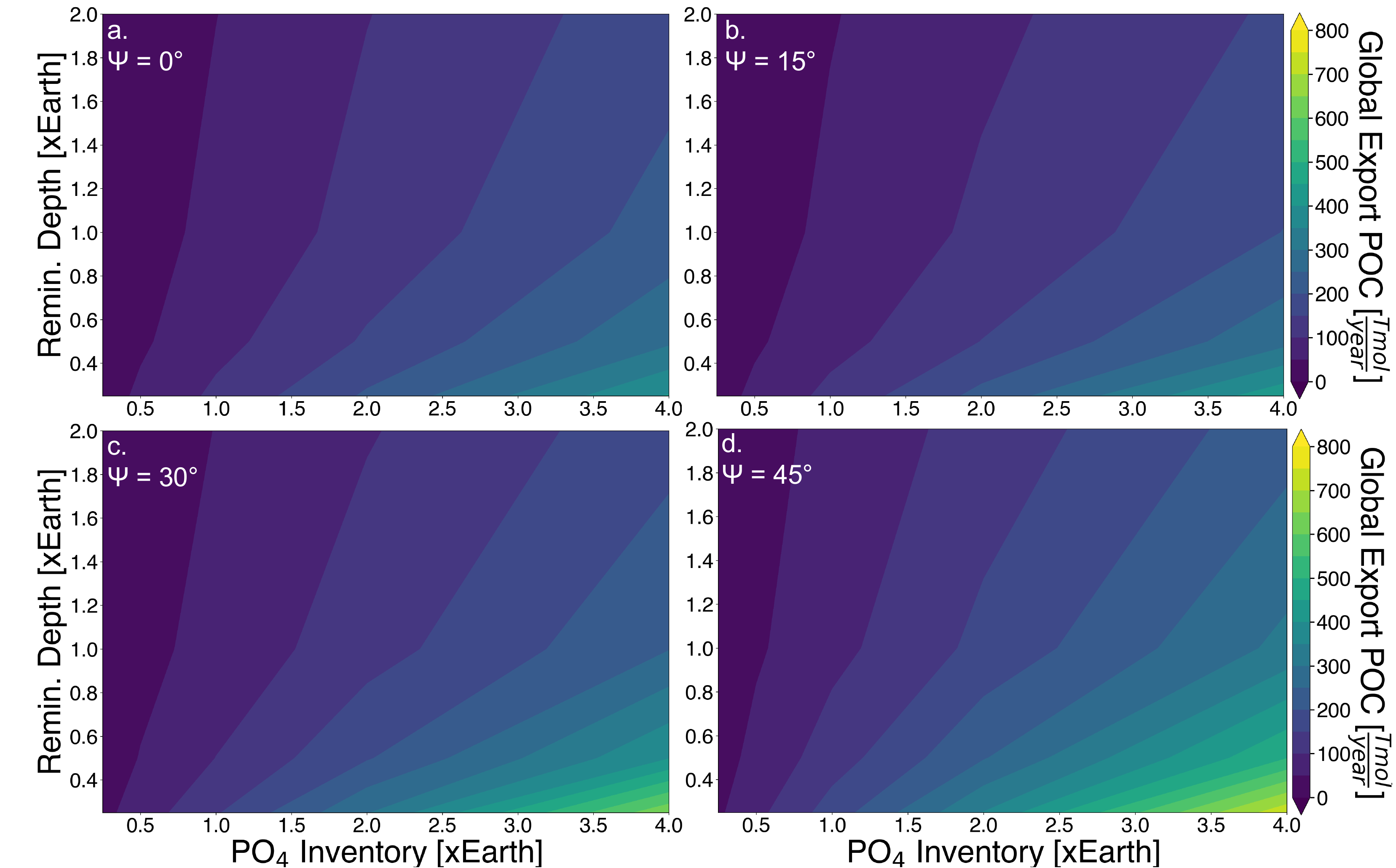}
    \caption{Global annual export particulate organic carbon for simulations with covaried phosphate inventory, remineralization length-scale, and obliquity. We consider obliquities of 0$\degree$ (panel a), 15$\degree$ (panel b), 30$\degree$(panel c), and 45$\degree$ (panel d). Annual global export POC increases with increasing phosphate inventory (\ref{fig:fig2}, decreasing remineralization length-scale (\ref{fig:fig3}), and increasing obliquity.}
    \label{fig:fig4}
\end{figure*}

\subsection{Additional Experiments}

We ran a series of experiments in which we covaried obliquity, remineralization length-scale, and phosphate inventory to explore potential synergistic effects introduced through covariance of these variables. Such effects could have profound implications for the potential for oxygen build-up on Earth-sized exoplanets.

\begin{figure*}
    \centering
    \includegraphics[width=\textwidth]{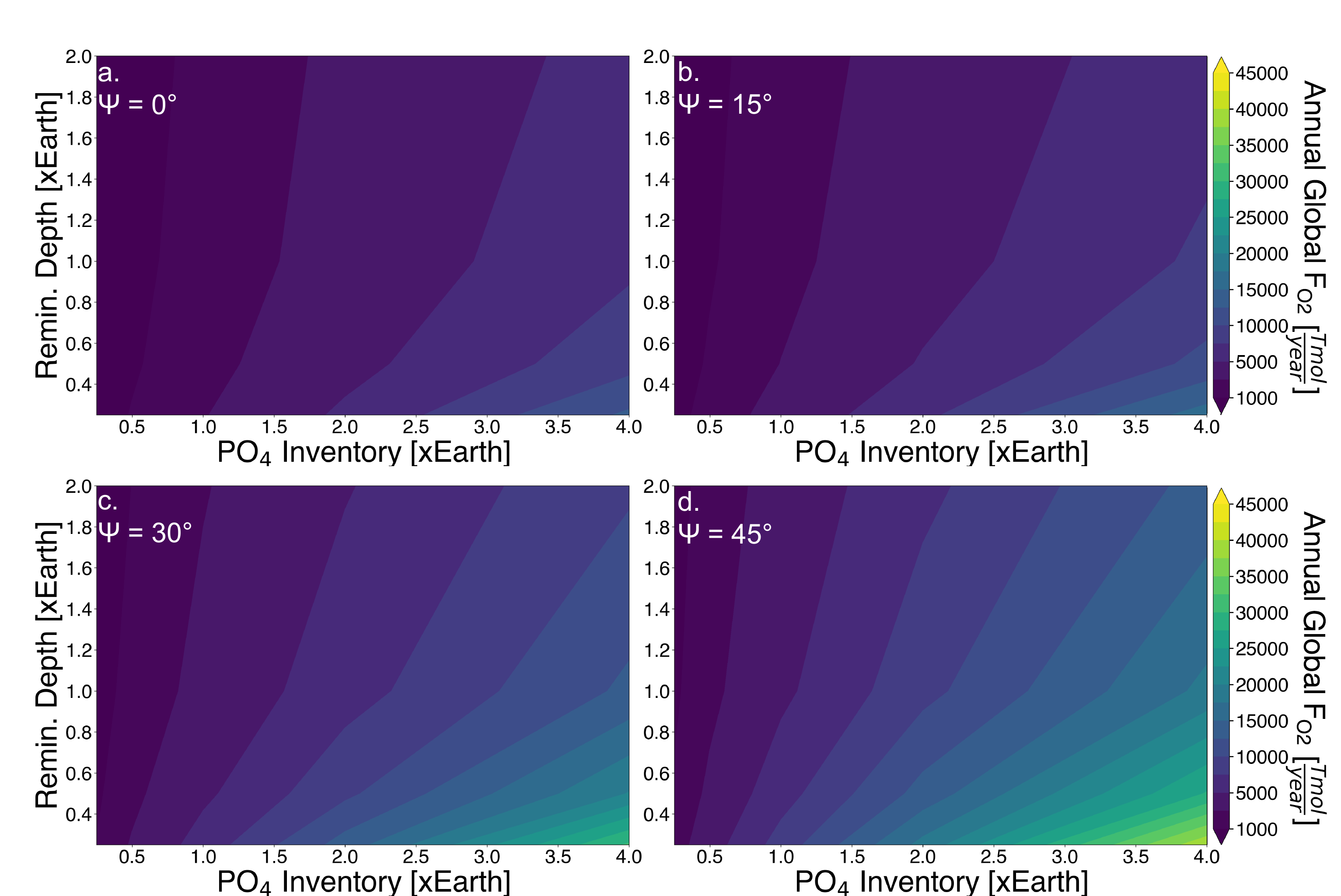}
    \caption{Annual global F$_{O_{2}}$ for simulations with covaried phosphate inventory, remineralization length-scale, and obliquity. We consider obliquities of 0$\degree$ (panel a), 15$\degree$ (panel b), 30$\degree$ (panel c), and 45$\degree$ (panel d). As seen in the single variable models, annual global F$_{O_{2}}$ increases with increasing phosphate inventory, decreasing remineralization length-scale, and increasing obliquity.}
    \label{fig:fig5}
\end{figure*}

\begin{figure*}
    \centering
    \includegraphics[width=\textwidth]{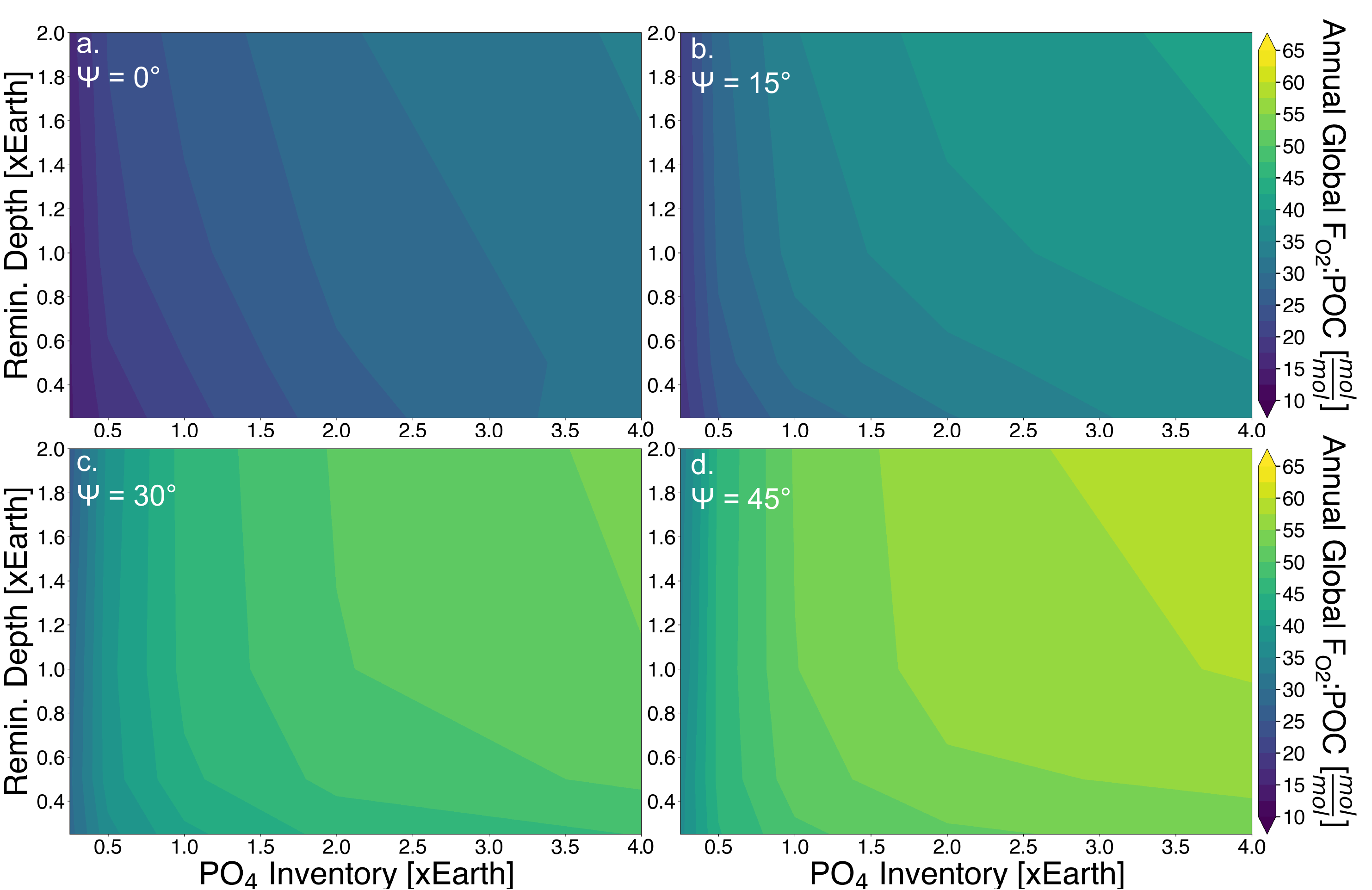}
    \caption{Annual global F$_{O_{2}}$ per unit POC for simulations with covaried phosphate inventory, remineralization length-scale, and obliquity. We consider obliquities of 0$\degree$ (panel a), 15$\degree$ (panel b), 30$\degree$(panel c), and 45$\degree$ (panel d). Global average F$_{O_{2}}$:POC increases with increasing phosphate inventory, increasing remineralization length-scale, and increasing obliquity. }
    \label{fig:fig6}
\end{figure*}

The general trends for export POC, F$_{O_{2}}$, and F$_{O_{2}}$:POC ratio in response to increasing obliquity, phosphate inventory, and remineralization length-scale previously discussed in sections 3.1, 3.2, and 3.3 continue to hold true in the co-varying simulations (Figures \ref{fig:fig4}, \ref{fig:fig5}, \ref{fig:fig6}). Both export POC and F$_{O_{2}}$ increase most strongly in response to increasing phosphate inventory, with second order increases driven by decreasing remineralization length-scale and increasing obliquity. When we increase phosphate inventory and decrease remineralization length-scale in tandem, a synergistic effect is present. We find that scenarios with increased phosphate inventory and decreased remineralization depth yield \textgreater2x export POC and F$_{O_{2}}$ levels than the sum of two independent scenarios with increased phosphate inventory and decreased remineralization length-scale.

\section{Discussion}
\label{sec:Discussion}
\subsection{Increased Obliquity Drives Enhanced Biological Activity and Atmospheric Oxygenation}

\begin{figure*}
    \centering
    \includegraphics[width=\textwidth]{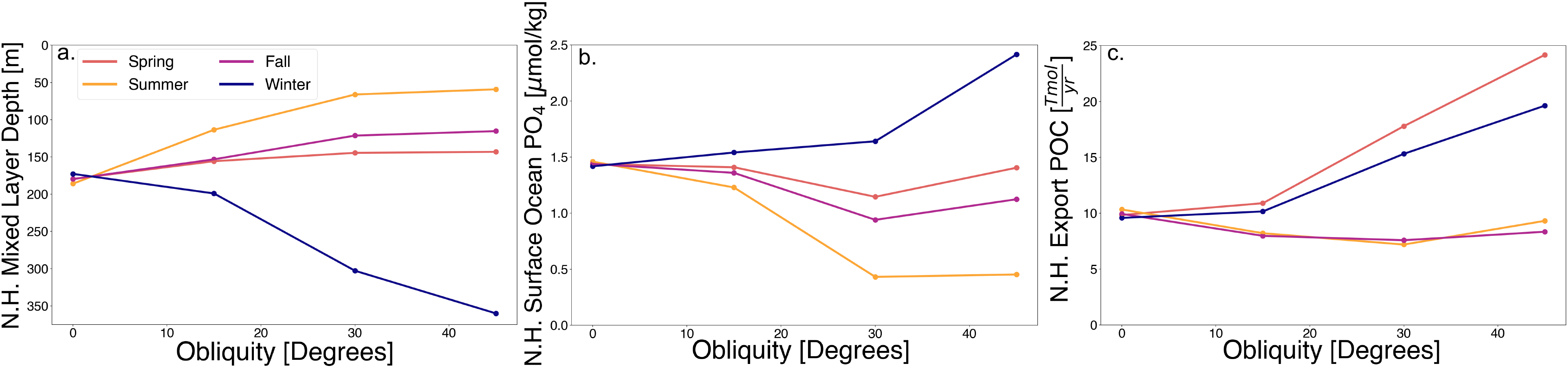}
    \caption{Seasonal Northern Hemisphere (N.H.) values for average mixed layer depth (panel a), total surface ocean PO$_{4}$ (panel b), and total export POC (panel c).}
    \label{fig:fig7}
\end{figure*}

The general increase in export POC with higher obliquity is driven by increased nutrient availability. As discussed earlier, \cite{olson_oceanographic_2020} predicted that increased seasonal deepening of the mixed layer depth with increasing obliquity during each hemisphere's winter would lead to increased nutrient cycling. This increased seasonal deepening of the mixed layer depth occurs with increasing obliquity in our scenarios as well (Figure \ref{fig:fig7}a). 

Seasonal deepening of mixed layer depth in the winter hemisphere intensifies with increasing obliquity because as obliquity increases, the winter hemisphere receives less incident insolation. Reduced incident insolation leads to colder sea surface temperatures and weakened water density stratification, resulting in a deeper mixed layer depth. Deeper mixed layer depths allow for increased entrainment of nutrients from depth, resulting in higher surface phosphate concentrations (Figure \ref{fig:fig7}b).

Varying seasonal nutrient availability significantly affects biological primary productivity throughout the year. In the 0$\degree$ obliquity scenario, POC export rates do not significantly vary throughout the year due to the lack of seasons (the slight variation between seasons displayed is due to the planet's non-zero eccentricity). However, for obliquity scenarios of 15$\degree$ or higher, spring has the highest rate of POC export, followed by winter (Figure \ref{fig:fig7}c). In the 45$\degree$ obliquity scenario, spring export POC increases by approximately 2.5x in comparison to the 0$\degree$ obliquity scenario, while winter export POC increases by a factor of two when compared to the 0$\degree$ obliquity scenario. In the non-zero degree obliquity scenarios, fall and summer have the lowest rates of POC export with summer POC export at slightly higher rates than in the fall.

POC export in the winter exceeds POC export in summer and fall due to enhanced nutrient inventories in the surface ocean (Figure \ref{fig:fig7}c). These nutrient rich waters are dredged up by the deepened mixed layer in these moderately high obliquity scenarios. However, although winter surface waters have an enhanced nutrient inventory, these nutrients cannot be entirely utilized due to reduced insolation during the winter. This inhibits biological activity, and therefore limits POC export until the spring. When winter ends, a ``spring bloom" of heightened biological activity occurs when incident insolation begins to increase again after the dark winter season and provides residual nutrient rich waters with light. Biological activity increases dramatically, fueled by both sufficient light and abundant nutrients, and leads to the observed maximum POC export in spring (Figure \ref{fig:fig7}c). As a result of the deepened mixed layer and enhanced winter nutrient availability, the increases in seasonal export POC for moderately high obliquity scenarios lead to the observed increase in annual export POC with obliquity.

Though light and temperature do have significant effects on POC export in general, in our simulations these effects are negligible compared to the effect of nutrient availability. As obliquity increases, light is more evenly distributed over the planetary surface on annual average but the total amount of light the planet's surface receives does not change and so light cannot be the cause of the observed increased annual export POC. Moderately high obliquity increases the amount of incident irradiation on a hemisphere during the summer months, and decreases it during the winter months. One might then expect an increase in export POC during the summer, and a decrease in the winter, which is not observed in our results due to the overriding effect of nutrient availability. Additionally, in cGENIE's temperature-dependent export scheme, export POC increases with increasing temperature. One may then expect that larger temperature extremes due to increased seasonality would lead to increased export POC in the summer and decreased export POC in the winter. However, as we instead see that winter export POC surpasses summer export POC (Figure \ref{fig:fig7}c), we conclude that nutrient availability is the dominant factor affecting export POC.

We find that both export POC and sea-to-air oxygen flux increase with increasing obliquity (Figures \ref{fig:fig2}, \ref{fig:fig3}, \ref{fig:fig4}, and \ref{fig:fig5}). This increase in F$_{O_{2}}$ is a direct consequence of increased biological activity; as biological activity increases there are more photosynthetic organisms performing photosynthesis, resulting in an increase in biogenic oxygen production. An interesting consequence of increased obliquity, however, is the effect of increasing obliquity on annual global oxygen flux per unit export POC. This increase implies that increased oxygen flux cannot be solely attributed to the increase in export POC.

If the increase in F$_{O_{2}}$ with increasing obliquity was simply due to the increase in export POC with obliquity, the F$_{O_{2}}$:POC ratio would be constant across all considered planetary obliquity values as produced oxygen would scale proportionally with increased primary productivity, which it is not (Figure \ref{fig:fig2}). An increase in F$_{O_{2}}$:POC implies that physical effects are likely amplifying the increase in F$_{O_{2}}$ with increasing obliquity.

\begin{figure*}
    \centering
    \includegraphics[width=\textwidth]{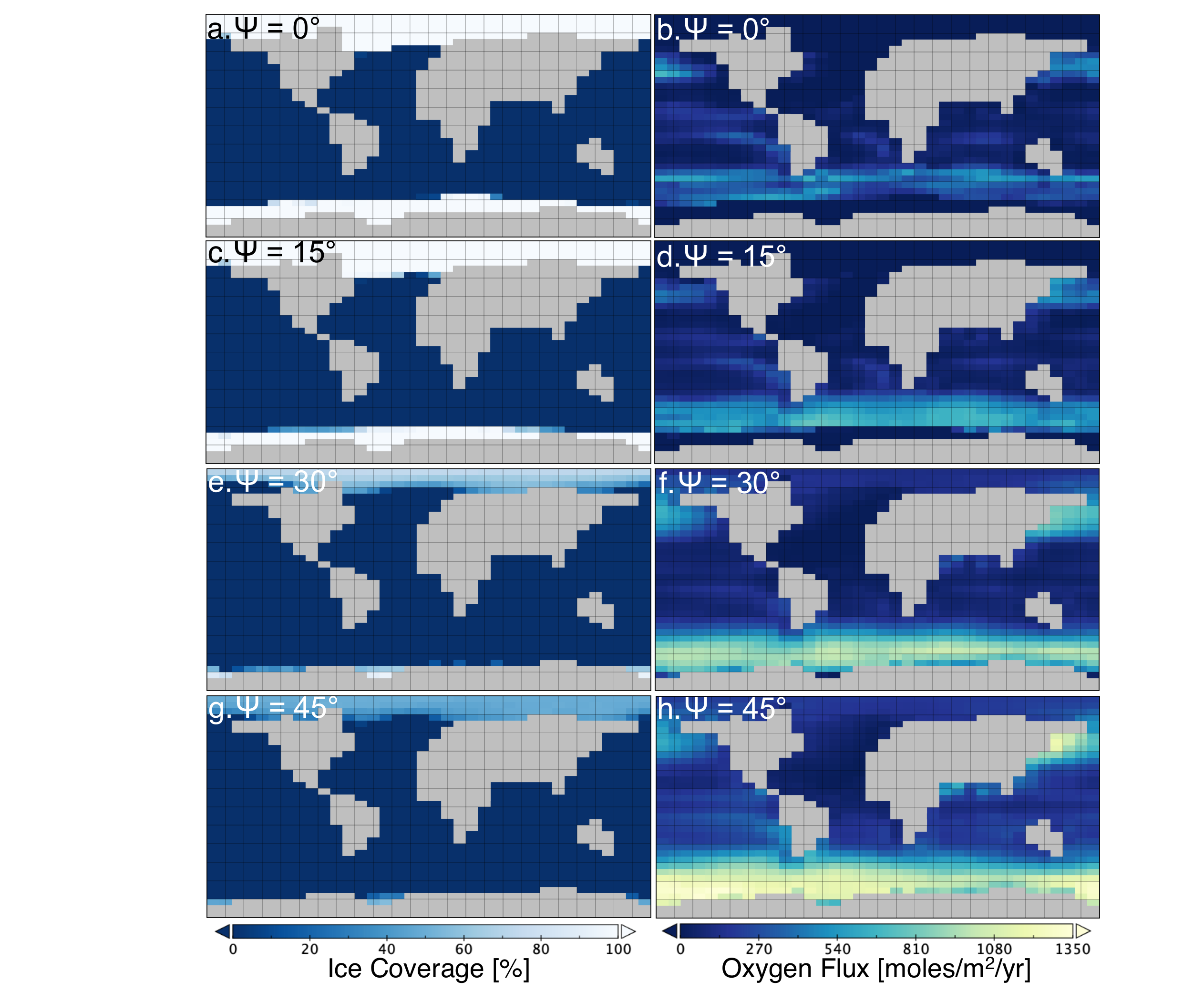}
    \caption{Global maps of annual average sea ice cover percentage and oxygen flux for simulations with POL phosphate inventory and remineralization length-scale. We consider obliquities of 0$\degree$ (panels a/b), 15$\degree$ (panels c/d), 30$\degree$(panels e/f), and 45$\degree$ (panels g/h). The left color bar denotes the percentage sea ice cover in each area box, while the right colorbar denotes annual average oxygen flux from each quadrant. The grey regions correspond to areas of land coverage, where ice coverage and oxygen flux are not measured.}
    \label{fig:fig8}
\end{figure*}

The physical effects responsible for amplifying F$_{O_{2}}$ increase with higher obliquity are decreased sea surface ice coverage and decreased solubility of oxygen in warmer water. Higher obliquity results in a more even distribution of insolation over the planetary surface. This leads to a decrease in the global average percentage of sea ice coverage as obliquity increases (Figure \ref{fig:fig8} a,c,e,g) and warmer waters due to higher incident irradiation at the poles. As the sea ice coverage decreases due to the increase in obliquity and increase in ocean water temperature, a corresponding increase in oxygen flux from those now uncovered areas at high latitudes occurs (Figure \ref{fig:fig8} b,d,f,h). The resulting increase in oxygen flux occurs because (1) without ice coverage, these areas of open ocean are now able to exchange oxygen with the atmosphere with no physical barrier impeding gas exchange and (2) oxygen is less soluble in warmer water \citep[e.g.][]{wanninkhof_relationship_2014}. Therefore, higher photosynthesis rates at the poles leads to enhanced sea-to-air fluxes of biogenic oxygen due to the combined effects of increased surface area available for gas exchange and warmer ocean temperatures.

\subsection{Moderately High Obliquity Increases Biosphere Detectability}

Atmospheric oxygen may be a remotely detectable signature of life on other planets \citep[e.g.][]{meadows_exoplanet_2018}. Although current telescopes are not designed to detect biotic levels of oxygen in exoplanet atmospheres \citep[][]{lustig-yaeger_detectability_2019}, next generation telescopes such as the LUVOIR-like observatory recommended by the Astro2020 Decadal Survey would be able to detect biotic levels of oxygen in exoplanet atmospheres depending on cloud conditions and exposure times \citep[][]{wang_baseline_2018}. As we have shown in our experiments, moderately high-obliquity planets have higher annual global oxygen fluxes from sea to air than their low obliquity equivalents (Figures \ref{fig:fig2}b, \ref{fig:fig3}b, \ref{fig:fig5}a-d). Therefore, biospheres on moderately high-obliquity exoplanets may be easier to detect than on their equivalent low obliquity counterparts due to their higher atmospheric oxygenation potential. 

Oxygen is not the only potential biosignature that is enhanced by moderately high obliquity. \citet{olson_atmospheric_2018} explored the potential for the use of seasonality as a biosignature, in which the seasonal increase and decrease of biogenic gases such as CO$_{2}$, CH$_{4}$, O$_{2}$, and O$_{3}$ due to seasonal variation in insolation could be detectable by a LUVOIR-like instrument and indicate the presence of an active biosphere. As we have seen here on Earth, CO$_{2}$ and O$_{2}$ oscillate seasonally due to changes in the balance of photosynthesis (draw down CO$_{2}$, make O$_{2}$) and respiration (release CO$_{2}$, consume O$_{2}$) \citep[][]{keeling_exchanges_2001}. While Earth has relatively moderate seasonality, exoplanets with higher obliquities would have more intense seasonality with large changes in temperature. As seasonal changes in the environment manifest as changes in atmospheric composition, larger seasonal differences may result in a larger, more detectable atmospheric signal \citep[][]{olson_atmospheric_2018}.

\subsection{Challenges for Life on Moderately High-Obliquity planets}

While aquatic life is protected from seasonality due to the high heat capacity of water, land environments may experience extreme seasonality on moderately high-obliquity planets. This could be a problem because terrestrial microbes and ectothermic organisms are unable to regulate their own body temperatures \citep[e.g.][]{wright_temperature_1981,cossins_temperature_1987, huey_evolution_1989, hochachka_biochemical_2016} and therefore would be unlikely to survive the temperature extremes on land resulting from increased seasonality \citep[similar to predictions by  ][ of exotherm response to temperature variations caused by climate change on Earth]{paaijmans_temperature_2013}. Additionally, while evolved terrestrial life that self-regulates body temperature (endotherms) can survive more extreme temperatures than ectotherms \citep[e.g.][]{scholander_adaptation_1950, irving_body_1954, heinrich_why_1977, lovegrove_perspectives_1991, mcnab_physiological_2002}, many adaptations to either extremely cold or hot environments are incompatible, such as fur or blubber.

Intensified seasonality does not have to be a death sentence for terrestrial life on moderately high-obliquity planets, however. Multiple microbial species and a few multi-cellular species have been discovered living in extreme environments on Earth, including those at extremely cold and hot temperatures \citep[e.g.][]{cavicchioli_extremophiles_2002}. Life evolving on a planet with extreme seasons could potentially adapt to both extreme warm and cold climates, and ultimately be more resilient against sudden environmental changes or mass extinctions.

Another potential problem for the origin and evolution of life on moderately high-obliquity planets is water loss. \citet{kang_wetter_2019} found that high-obliquity planets have a much higher seasonal stratospheric water vapor abundance in comparison to equivalent low obliquity planets, which may allow more water vapor to be lost due to photodissociation and hydrodynamic escape of hydrogen to space. This potential for increased water loss suggests that liquid water on high-obliquity planet surfaces could be lost quickly, potentially endangering the origin/evolution of life on these planets as liquid water is a key requirement for life. However, \citet{kang_wetter_2019} was looking at planets with obliquities higher than those considered in this work, so moderately high-obliquity planet atmospheres may not experience the wetter stratospheres seen in higher-obliquity planet atmospheres. Additionally, potential water loss may be mitigated by the fact that wetter stratospheres only occurred seasonally in simulations from \citet{kang_wetter_2019}, and may average out to negligible increased water loss on long-term average for high-obliquity exoplanets and their moderately high-obliquity counterparts.

\subsection{Opportunities for Future Work}
Our results suggest that moderately high-obliquity planets may experience both higher export POC production and higher oxygenation potential than their low-obliquity counterparts. However, as we only simulate planetary obliquities up to 45$\degree$, it is still unknown how the biospheres on higher-obliquity planets will fare under the physical and climatic changes that occur at higher obliquities. At planetary obliquities greater than 54$\degree$, the planet enters a different climate regime where the poles begin to receive more incident insolation than the equatorial regions, potentially leading to the formation of equatorial ice belts \citep[e.g.][]{rose_ice_2017, kilic_multiple_2017, kilic_stable_2018, colose_enhanced_2019}. These equatorial ice belts would cause changes to both ocean circulation and surface wind fields, which may have significant effects on biospheric activity. We did not explore these obliquity scenarios because imposing annually averaged winds from ROCKE-3D would likely become an increasingly poor assumption at very high obliquities. Modeling biospheric dynamics and oxygenation potential on planets with higher obliquities remains an opportunity for future work.

\section{Conclusions}
\label{sec:Conclusions}
We simulated Earth-like marine life on an Earth-sized planet at various obliquities. We found that moderately high obliquity promotes increased photosynthetic activity and associated oxygen flux to the atmosphere, potentially enhancing atmospheric oxygenation. Sea-to-air oxygen fluxes are further amplified by decreasing sea surface ice coverage at high latitudes on moderately high-obliquity planets, exposing greater ocean surface area for sea-to-air gas exchange. This oxygenation may ultimately be beneficial for the evolution of complex life on moderately high-obliquity planets \citep[e.g.][]{catling_why_2005, reinhard_earths_2016}. In short, our results suggest that life may not just survive, but also thrive at moderately high obliquity.

Active biospheres may be easier to detect on moderately high-obliquity planets as well. Increased planetary oxygenation on moderately high-obliquity planets could result in easier detection of biogenic oxygen on moderately high-obliquity planets with next generation telescope concepts and potential earlier evolution of complex life \citep[e.g.][]{catling_why_2005, reinhard_earths_2016}. Additionally, life on moderately high-obliquity planets may be easier to detect through heightened variations in biosignature gases over more intense seasonal cycles. Given the potential for increased biological activity and oxygenation on moderately high-obliquity planets, the habitability and biosignatures of moderately high-obliquity planets are exciting opportunities for future work.

\bibliography{references_updated}

\end{document}